\documentclass[superscriptaddress,showpacs,twocolumn,amssymb,aps]{revtex4-1}

\usepackage{graphicx,dcolumn,bm,amsmath,color,bm}

\newcommand{\eq}[1]{Eq.~(\ref{#1})}

\newcommand{\fig}[1]{Fig.~\ref{#1}}

\newcommand{\eeq}{ \end{equation} }
\newcommand{\beq}{ \begin{equation} }

\newcommand{\eea}{ \end{align} }
\newcommand{\bea}{ \begin{align} }

\newcommand{\bhu}{ {\bf \hat{u}} }
\newcommand{\bhe}{ {\bf \hat{e}} }

\newcommand{\bfr}{ {\bf r} }

\newcommand{\be}{ {\bf \hat{e}} }

\newcommand{\bx}{ {\bf \hat{x}} }

\newcommand{\bv}{ {\bf \hat{v}} }
\newcommand{\bw}{ {\bf \hat{w}} }
\newcommand{\bn}{ {\bf \hat{n}} }

\begin{document}

\title{Chiral assembly of weakly curled hard rods: effect of steric chirality and polarity}
\author{H. H. Wensink}
\email{wensink@lps.u-psud.fr}
\author{L. Morales-Anda}
\affiliation{Laboratoire de Physique des Solides - UMR 8502, Universit\'e Paris-Sud  \& CNRS, 91405 Orsay, France}

\date{\today}

\begin{abstract}

We theoretically investigate the pitch of lyotropic cholesteric phases composed of slender rods with steric chirality transmitted via a weak helical deformation of the backbone.  In this limit, the model is amenable to analytical treatment within Onsager theory and a closed expression for the pitch versus concentration and helical shape can be derived.  Within the same framework we also briefly review the possibility of alternative types of chiral order, such as twist-bend  or  screw-like nematic phases, finding that cholesteric order dominates for weakly helical distortions.
While long-ranged or `soft' chiral forces usually lead to a pitch decreasing linearly with concentration, steric chirality leads to a much steeper decrease of quadratic nature. This reveals a subtle link between the range of chiral intermolecular interaction and the pitch sensitivity with concentration.  A much richer dependence on the thermodynamic state is revealed for polar helices where parallel and anti-parallel pair alignments along the local director are no longer equivalent.  It is found that weak temperature variations may lead to dramatic changes in the pitch, despite the lyotropic nature of the assembly.  

\end{abstract}

\maketitle

\section{Introduction}
Chirality is ubiquitous in the natural world and plays an important role 
in biological systems. Manifestations of chiral symmetry are evident on 
the molecular scale (e.g. the double helix structure of DNA) as well on 
the macroscopic scale (think of e.g. spiral patterns in snail shells and 
plant morphologies). It is also prominent in condensed 
phases of anisometric building blocks  forming  liquid crystal 
mesophases \cite{gennes-prost}. Well-known examples range from cholesteric liquid crystals 
whose helical meso-structure is of key importance for use in liquid 
crystal display technology to more exotic chiral mesostructures found in bent-core or banana-shaped molecules \cite{jaklibentcore2013}.

The main motivation for studying chiral assemblies lies in the fact that many biomacromolecules have a helical internal structure which plays an important role in the structure and funciontality of the cell.  Examples of chiral supramolecular organization can be found in assemblies of helical polymers such as actin \cite{claessensPNAS2008}, microtubules \cite{huang2012}, DNA \cite{livolantDNAoverview, kornyshevRMP2007}, filamentous virus particles  \cite{dogicfradenLA2000} and bacterial flagella \cite{barryPRL2006}.
Other ways to deepen our understanding of chiral self-assembly is by studying artificial macromolecules with a helical signature such as carbon nanotubes \cite{amelinckx}, cellulose nanocrystals \cite{lagerwall2014,schutz-cellulose2015} or helical polymer-nanoparticle ribbons \cite{crosbyAM2013}.

Theoretical attempts to predict the behavior of the chiral mesostructure of condensed phases have met with variable success. Formulating a sound statistical mechanical theory of these systems is extremely challenging in view of the complexity of the underlying chiral interactions, the anisotropic shape of the building blocks and the inhomogeneous and anisotropic symmetry of the structure \cite{harris-kamienPRL1997,emelyanenkoPRE2003}. 
As far as cholesteric order goes, most theoretical predictions for the pitch modulation based on microscopic theory to date have focussed either on long-ranged chiral forces of a dispersive nature treated within (van der Waals) perturbation theory \cite{goossens1971,odijkchiral,vargachiral1,wensinkEPL2014,wensinkJPCM2011} or on hard helices of arbitrary amplitude and pitch where elaborate numerical schemes are required to compute the intricate excluded-volume terms in the free energy \cite{belliPRE2014,kolliSM2014}.  
Other mechanical models for steric chirality that have been proposed are based on density functional theory of twisted hard boards \cite{evansMP1992}, Maier-Saupe theory of rigid and semiflexible corkscrews \cite{pelcovits1996} and Onsager-Parsons theory for chiral two-site segment particles with planar orientations \cite{vargajackson2011}. Unfortunately, the predictions ensuing from these models  do not agree as to the concentration dependence of the pitch. Amongst these steric models only the corkscrew model, developed by Pelcovits \cite{pelcovits1996}, predicts a marked dependence on concentration as observed in experiment \cite{dupreduke75,dogicfradenLA2000,gray-cullulose,miller-cellulose,schutz-cellulose2015}.  

The focus of this study is to revisit the pitch of weakly curled helices, a regime that is relevant to many helical biomolecules, by seeking qualitative guidance from Onsager-Straley theory \cite{onsager,straleychiral}. The main motivation for using a second-virial theory resides in the fact that the excluded volume of weakly deformed helices can be determined analytically thus greatly expediting the calculation of the elastic and  torque-field parameters that govern the helical mesostructure.  In addition, we wish to broaden the context of the theory beyond mere cholesteric order by attempting to gauge the possibility of alternative types of chiral mesophases such as the twist-bend \cite{dozovEPL2001,memmer2002,borschNC2013} and screw-like nematics reported for non-convex elongated particle shapes \cite{kolliJCP2014}.

The objective of our theoretical study is threefold.  The first is to ascertain which type of helical mesostructure (cholesteric, twist-bend or screw-like nematic) prevails for particles with weak steric chirality.  Second, we wish to establish tractable expressions linking the pitch to the helix concentration and shape and make a comparison with previous theoretical predictions and experimental observations. Last not least, we aim to take a closer look at the effect of weak polar forces between neighbouring helices and scrutinize the effect of spontaneous polarization along the local director on the pitch. Many helical nanoparticles are intrinsically polar due to their distinct head-tail asymmetry. In this light, our results could give useful guidance for the interpretation of mesoscale chiral order in suspensions of chiral biomacromolecules or artificial helical mesogens and offer routes to manipulating their helical mesostructure through subtle variations of particle concentration and temperature.

The rest of the paper is structured as follows. We begin by deriving a general expression for the excluded volume of inifinitely slender, weakly curled hard rods (Section II).  This quantity serves as a basis for a statistical mechanical theory from which we can estimate the stability of various types of chiral nematic order (cholesteric in Section III and twist-bend order in Section IV) as well as make quantitative predictions for the pitch in relation to the particle shape and the thermodynamical state of the system.  In Section V, we generalize the original mean-field theory  towards weakly polar interactions between helix pairs and scrutinize their impact on the cholesteric pitch. We end by formulating the basic conclusions, as well as some avenues for future research.

\section{Excluded volume of weakly curled rods}

\begin{figure}
\begin{center}
\includegraphics[width= \columnwidth]{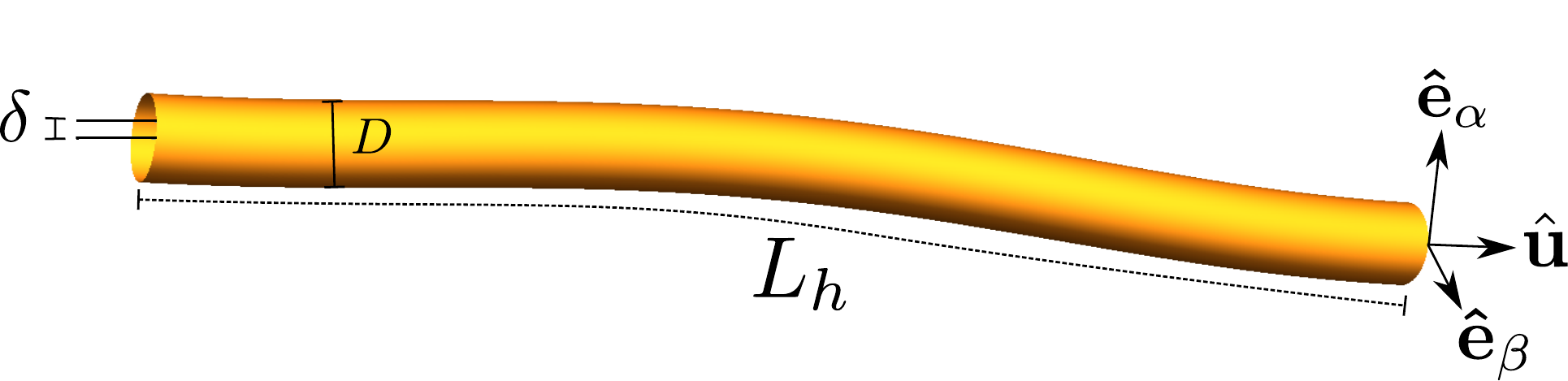}
\caption{Cartoon of a weakly curled rigid hard rod with helical amplitude $\delta$ and its corresponding particle frame $\{ \bhu, \bhe_{\alpha}, \bhe_{\beta} \}$. The contour length $L_{h} \approx L(1+ \frac{1}{2} (q \delta)^{2})$ is defined in terms of the amplitude, projected length $L$ and inverse molecular pitch $q$. The aspect ratio is not to scale and should respect the Onsager limit $L_{h}/D \rightarrow \infty$. }
 \label{fig0} 
\end{center}
\end{figure}

Consider a thin helical rod with  length $L$ and thickness $D$ in the Onsager limit of infinite aspect ratio $L/D \gg 1 $. The backbone is described by the following vector parameterizing a helix with principal orientation unit vector $\bhu$ 
\beq
\bfr_{h}(t)  = \bfr_{0} + \frac{tL}{2}  \bhu  +  \frac{\delta}{2} ( \be_{\alpha}  \cos qL t   + \be_{\beta}  \sin qL t )  
\label{parhel}
\eeq
with  $-1<t<1$, $\delta \ll L$ the helix radius assumed small, and $q = 2\pi  / p$ a wavenumber inversely proportional to the the pitch length $p$. Two auxiliary unit vectors are introduced so that $\{ \bhu, \be_{\alpha}, \be_{\beta} \}$ represents a particle-based orthonormal frame (see \fig{fig0}). 
In case of a pair of helices with main directions $\bhu_{1}$ and $\bhu_{2}$ it is advantageous to define additional unit vectors $\bv = \bhu_{1} \times \bhu_{2}/ | \bhu_{1} \times \bhu_{2} | $ and $\bw_{i} = \bhu_{i} \times \bv $ ($i=1,2$) so that $\{\bhu_{i} , \bv , \bw_{i} \}$ constitutes two orthonormal molecular frames.  We will therefore identify $\be_{\alpha1} = \be_{\alpha2} = \bv  $ and $ \be_{\beta i} = \bw_{i}$. 
The centre-of-mass distance describing the overlap volume between two helices with equal shape but different orientation bases is then given by
\begin{align}
\Delta \bfr_{cc} & =  \frac{L}{2} (t_{2} \bhu_{2}  -t_{1} \bhu_{1} )    + \frac{\delta }{2} ( c_{2}  -  c_{1}) \bv  + \frac{\delta  }{2}  (\bw_{2}s_{2}  -  \bw_{1} s_{1}) \nonumber \\ 
& + D \left  ( 1 + \frac{(q \delta)^{2}}{16} (s_{1}^{2}  + s_{2}^{2} )  \right ) t_{3} \bv
\label{pm}
\end{align}
with  $-1 < t_{i} < 1$ ($i =1,2,3 $) and introducing short-hand notation
$s_{i}  =  \sin (q L t_{i} + \psi_{i}) $ and $c_{i}  = \cos (q L t_{i} + \psi_{i})$. 
The last term in \eq{pm} contains curvature-dependent contributions of which we have only retained the lowest order quadratic contributions which should suffice for $q \delta < 1$.

Since a helix is a no longer a uniaxial object all pair correlations are variant upon rotations of the helices around their main orientation axis.  These rotations in the $\{ \bv, \bw_{i}\} $-plane perpendicular to the main orientation $\bhu$ are described by a rotation matrix ${\mathcal R}(\psi)$ with internal angle $\psi$, so that
$ {\mathcal R}(\psi_{i}) \cdot (\sin (q L t_{i}) , \cos(q L t_{i})  = (s_{i}, c_{i} )$.  We now proceed to calculate the excluded volume between a pair of particles defined as
\begin{align}
v_{\rm excl}    & = \int_{-1}^{1} dt_{1} \int_{-1}^{1} dt_{2} \int_{-1}^{1} dt_{3} | {\bf J} |
\end{align}
in terms of the Jacobian matrix $J_{m n}  = \partial \Delta r_{cc,m}/ \partial t_{n} $ with $ m = x,y,z$ and $ n = 1,2,3$. The Jacobian determinant can be calculated analytically and reads up to quadratic order in $q \delta$
\begin{align}
\frac{| {\bf J} |}{ \frac{1}{4} L^{2}D } & =    q \delta  (c_{1} - c_{2}) | \cos \gamma | \nonumber \\ 
& + \left \{ 1  + (q \delta )^{2} (c_{1} c_{2} + \frac{1}{16}(s_{1}^{2} + s_{2}^{2}) \right \}  | \sin \gamma | 
\end{align}
 with $\gamma$ the angle between the two main orientation vectors. The excluded volume turns out to have the following form
 \begin{align}
& \frac{v_{\rm excl} (\gamma, \psi_{1}, \psi_{2})} { 2 L^{2} D}   =  q \delta  (\cos \psi_{1} - \cos \psi_{2})  |\cos \gamma | + | \sin \gamma | \nonumber \\
& \times \left (1 + (q \delta)^{2} [ \cos \psi_{1} \cos \psi_{2} + \frac{1}{32}(2 - \cos2 \psi_{1} - \cos 2 \psi_{2} )]   \right )    \nonumber \\  \label{vex1}
 \end{align}
 which reduces to the well-known expression $v_{\rm excl}(\gamma) = 2L^{2}D | \sin \gamma |$ for straight rods ($\delta =0$). 
  A simple configuration with `zero measure' is the one where both particles have zero internal angle $\psi_{1} = \psi_{2}=0$. \eq{vex1}  then simply reproduces the result for a straight rod 
 \begin{align}
 \left . v_{\rm excl} (\gamma)  \right |_{\psi_{i}  =0} &=   2 L^{2} D \left (1 +   (q \delta)^{2} \right ) | \sin \gamma | \nonumber \\ 
 & = 2 L_{h}^{2} D |\sin \gamma |    
 \end{align} 
 with the bare rod length $L$ replaced by the helix contour length $L_{h}/L \approx 1 + \frac{1}{2}(q \delta)^{2} $.
 \eq{vex1} can be  rendered more insightful by  defining the relative internal angles $\bar{\psi} = ( \psi_{1} + \psi_{2})/2 $ and $\Delta \psi = \psi_{2} - \psi_{1}$ so that $\psi_{1/2} = \bar{\psi} \mp \Delta \psi $.  The angle $\bar{\psi}$ between the azimuthal direction of the helix and the vector $\bv$ in the particle frame may assume a random value. This leads to an expression depending on the relative angles only
 \begin{align}
 \frac{v_{\rm excl} (\gamma, \Delta \psi)} { 2 L^{2} D  | \sin \gamma | }   = 1 +   (q \delta)^{2} \left ( \frac{1}{16} +  \frac{1}{2} \cos \Delta \psi   \right )    \nonumber \\  \label{vex2}
 \end{align} 
This result is, as expected, independent of the helix handedness determined by the sign of $q$. We reiterate that, due to the parameterization chosen in \eq{pm}, the above expression only gives a true representation of the excluded volume for weakly curled rods with $q\delta \ll  1$. We have also neglected end effects associated with finite aspect ratio $L/D$ which are at least of order ${\mathcal O}(LD^{2})$. 
In case the internal angles are distributed randomly over the interval $\Delta \psi \in [-\pi,  \pi] $ the excluded volume becomes
 \beq
 \langle \langle  v_{\rm excl} (\gamma) \rangle \rangle_{\psi}  =   2 L^{2} D \left (1 + \frac{1}{16} (q \delta)^{2} \right ) | \sin \gamma |  
 \eeq 
 from which we deduce, not surprisingly,  that  curling up a straight rod imparts an effective thickness $>D$ and an increase of the excluded volume.

\section{Isotropic-nematic bifurcation}
 
\begin{figure}
\begin{center}
\includegraphics[width= \columnwidth]{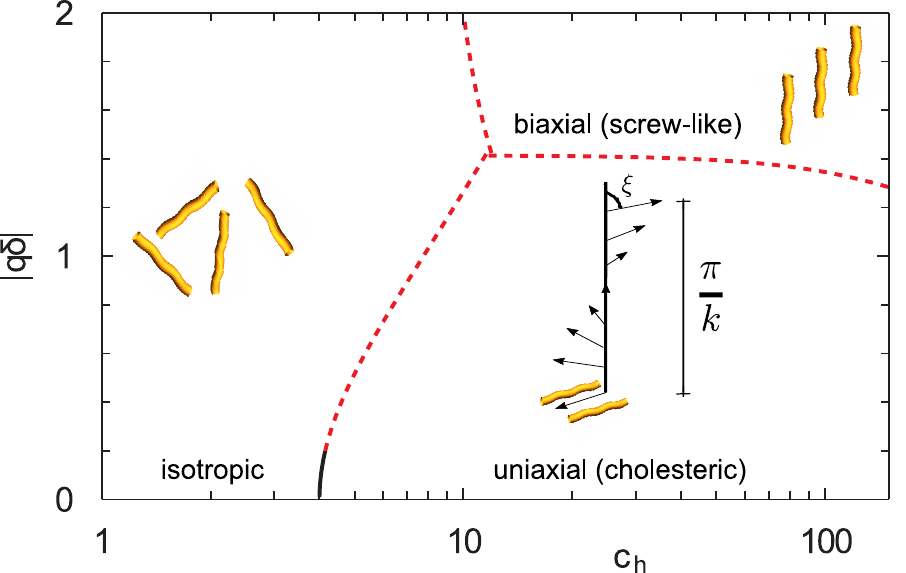}
\caption{ \label{fig1} Tentative phase diagram for hard helical rods roughly demarcating the stable phases upon increasing concentration $c_{h}$: isotropic, cholesteric and biaxial nematic.  For weakly curled helices $q \delta \ll 1$ a transition to the screw-like biaxial nematic is expected at very high rod concentration. The focus of the present theory is on the regime  $q \delta \ll 1$ indicated by the solid black line. }
\end{center}
\end{figure}

Onsager theory \cite{onsager,vroege92} dictates that the Helmholtz free energy $F$ in units of the thermal energy $k_{B}T$ of an ensemble of $N$ slender hard rods in a volume $V$ with number density $\rho=N/V$ is described by a simple second-virial form
 \begin{align}
\frac{F}{Nk_{B}T} &= \int d \omega f(\omega)  ( \ln [ 8 \pi^{2} f(\omega)] -1 ) \nonumber \\ 
& + \frac{\rho}{2} \iint d \omega_{1} d \omega_{2} f(\omega_{1})f(\omega_{2}) v_{\rm excl} (\omega_{1}, \omega_{2} ) 
\label{f0}
\end{align}
where the excluded volume between the particles under consideration serves as a key input. Analogous to  classical density functional theory the free energy involves an unknown orientational distribution function (ODF) $f$ describing the probability to find the main helix vector pointing at a certain solid angle $\Omega$ on the unit sphere with an internal angle $\psi$. For brevity we have introduced the shorthand notation $\omega = \{ \Omega, \psi \}$ to denote the total orientational phase space so that $d\omega= d \Omega d \psi$ and $\int d \omega = 8 \pi^{2}$.  Minimizing $F$ with respect to $f$ yields an Euler-Lagrange equation 
\beq
\ln f(\omega_{1}) + \rho \int d \omega_{2} f(\omega_{2}) v_{\rm excl} (\omega_{1}, \omega_{2} ) - \lambda \int d \omega f(\omega)=0 \label{el}
\eeq
with $\lambda$ a multiplier ensuring normalization of $f$. Let us now seek instabilities of the isotropic phase, in which all (internal) helix orientations are equally probable, to a nematic one.  The main helix orientation describes a unit sphere $d \Omega = \sin \theta d \theta d \varphi$ in terms of a polar angle $0<\theta < \pi $ and azimuthal one  $0 < \varphi < 2 \pi$. Including the internal angles it follows that $f$ in the isotropic phase corresponds to the inverse of the volume of the total orientational phase space $f(\omega) = 1/8\pi^{2}$. In order to probe transitions to nematic order we consider the following nematic perturbation
 \beq
 f(\omega) = \frac{1}{8 \pi^{2} } ( 1 + \varepsilon_{1} {\mathcal P }_{2}(\cos \theta) + \varepsilon_{2} {\mathcal P }_{2}^{1} (\cos \theta) \cos \varphi \cos \psi ) \label{perturb}
 \eeq
with ${\mathcal P}_{n}$ a standard Legendre polynomial and ${\mathcal P }_{n}^{m}$ an associated one. Here, $\varepsilon_{1}$ denotes an infinitesimally small amplitude perturbation towards a uniaxial nematic phase (with $\psi$ randomly distributed), whereas $\varepsilon_{2}$ corresponds to a biaxial mode where the distribution of the internal angles and the azimuthal one is no longer uniform.  The sine contribution in the excluded volume can be expanded as follows \cite{Lekkerkerker84}
 \beq
| \sin \gamma | = \frac{\pi}{4} + \sum_{n=1}^{\infty} d_{2n} {\mathcal P}_{2n} (\cos \gamma) 
 \eeq
with first coefficient $d_{2} = -5 \pi/32$.  A similar expansion exist for the cosine but this contribution turns out immaterial for the subsequent analysis.  Given that helices obey inversion symmetry, only even Legendre polynomials need be retained. By virtue of
the addition theorem of spherical harmonics  the functions ${\mathcal P}_{2n}(\cos \gamma )$ can be expressed as a bilinear expansion in the polar and azimuthal angle
\begin{align}
& {\mathcal P}_{2n}(\cos \gamma )  = {\mathcal P}_{2n}(\cos \theta_{1} ){\mathcal P}_{2n}(\cos \theta_{2}) \nonumber \\ 
& +2\sum_{m=1}^{2n}\frac{(2n-m)!}{(2n+m)!}{\mathcal P}_{2n}^{m}(\cos \theta_{1}){\mathcal P}_{2n}^{m}(\cos \theta_{2}) \cos  m \Delta \varphi  \label{3addition}
\end{align}
Bifurcations from the isotropic ODF can be sought by inserting \eq{perturb} into  \eq{el}, linearizing with respect to the amplitudes $\{ \varepsilon_{1}, \varepsilon_{2} \}$ and using the orthogonality properties of the (associated) Legendre polynomials. The density  $\rho^{\ast}$ corresponding to the isotropic-uniaxial nematic transition is 
\beq
\frac{\pi}{4} \rho^{\ast} D L^{2}   \simeq \frac{4}{(1 + \frac{1}{16} (q \delta)^{2} ) }
\eeq
It is, however, more appropriate to define a dimensionless Onsager density at a {\em fixed contour} length $L_{h}$ via $c_{h}= \frac{\pi}{4} L_{h}^{2} D \rho$ so that fixing $c_{h}$ preserves the total particle mass. The isotropic-uniaxial nematic bifurcation concentration up to quadratic order in curvature then becomes
\beq
c_{h}^{\ast}  \simeq 4 \left (1 + \frac{15}{16} (q \delta)^{2} \right )
\eeq  
reducing to the well-known result $c_{h}^{\ast} = 4$ for straight rods \cite{kayser}. We conclude that finite curliness shifts the isotropic-nematic transition to higher concentration due to a reduction of the (effective) aspect ratio of the object \cite{frezzaJCP2013}.  The result for the isotropic-biaxial nematic transition reads
\beq
c_{h}^{\ast} \simeq 8 \left ( 1+ \frac{1}{(q \delta)^{2}} \right )
\eeq
Clearly, in the weak curvature regime the  transition to a uniaxial nematic symmetry  always pre-empts the transition from an isotropic to a biaxial phase.  The stable incipient nematic phase  must therefore be of uniaxial symmetry with the main helix vectors pointing along a common director but with the internal direction distributed randomly. At higher density, however, the uniaxial phase might still crossover to a biaxial one. Such transitions can be probed by considering the following symmetry breaking perturbation to a nematic reference state with ODF $f_{0}(\theta)$
\beq
f(\omega) = f_{0}(\theta) + \varepsilon f_{1}(\theta) \cos \varphi \cos \psi 
\eeq 
We may now repeat the analysis above by inserting the above expression into the Euler-Lagrange equation and linearizing with respect to the amplitude $\varepsilon$.  Using the orthogonality property of the Legendre  and cosine functions it can be shown that the uniaxial-biaxial nematic transition follows from the condition
\beq
f_{1} (\theta_{1}) = - 2 c_{h} (q \delta)^{2} f_{0}(\theta_{1}) \int_{-1}^{1} d(\cos \theta_{2}) K_{1} (\theta_{1} , \theta_{2}) f_{1}(\theta_{2}) \label{eve}
\eeq 
with kernel 
\beq
K_{1} (\theta_{1} , \theta_{2} ) = \frac{4}{\pi} \iint_{0}^{2 \pi} \frac{d \varphi_{1}}{2 \pi}   \frac{d \varphi_{2}}{2 \pi} |\sin \gamma | \cos  \varphi_{1} \cos  \varphi_{2} 
\eeq
The uniaxial-biaxial nematic bifurcation concentration $(c_{h}^{\ast})_{UB}$ can be identified with the largest eigenvalue (corresponding to the eigenfunction $f_{1}(\theta)$) for a given reference state with $f_{0}(\theta)$ which itself is a solution of  the Euler-Lagrange equation \eq{el} for a given initial density. The required iteration procedure can be carried out numerically by discretizing the angular space into an equidistant grid. The results in \fig{fig1}  give an overview of the stability of the isotropic and nematic phases. The phase transition from isotropic to uniaxial nematic is of first order and coexistence densities (not calculated here) follows from equality of pressure and chemical potential which can be derived from the Helmholtz free energy. The uniaxial-biaxial nematic one turns out to be continuous.
The regime of weakly curled rods ($q \delta \ll 1 $) is dominated by uniaxial nematic order while biaxiality enters only at very high particle concentration. The high-density biaxial nematic phase has both chiral and polar symmetry and should be similar in nature to the screw-like nematic phase recently found in simulation \cite{kolliSM2014,kolliJCP2014} 
and previously observed in dense assemblies of helical flagella \cite{barryPRL2006}. In these dense structures the main helix axes align along a homogeneous (untwisted) director but their perpendicular direction (described by the vectors $\bv$ or $\bw$ in our case) follow a chiral precession with a pitch modulation corresponding to the microscopic pitch of the helices.  Although the theory is no longer quantitatively reliable beyond the regime of small curvature, roughly demarcated by the solid line in \fig{fig1}, we expect the bifurcation lines to provide the correct trends as to the onset of uniaxial and biaxial nematic order as a function of concentration and particle shape. The main purpose of this exercise is to underline the notion that  for weakly curled rods stable uniaxial order persists up to very high particle concentration
and that a very strong degree of polar alignment is needed to freeze the internal orientational degrees of freedom.

\section{Cholesteric structure}
In order to accommodate for  cholesteric order we invoke Straley's  approach \cite{straleychiral} to compute the free energy change associated with weak spatial distortions of the director field \cite{allenevans,odijkchiral}.  Let us define a spatially nonuniform director field $\bn ({\bf R})$. For weak spatial modulations of  $\bn ({\bf R})$ it suffices to consider the following double gradient contribution describing the elastic response to a weak deformation of the director field
\begin{align}
\frac{\Delta F_{2}}{k_{B}T} & = \frac{\rho^{2}}{4} \iint d {\bf R} d \bfr \iint d \bhu_{1} d \bhu_{2} \dot{f} (\bn ({\bf R}) \cdot \bhu_{1}) \dot{f} ((\bn ({\bf R}) \cdot \bhu_{2}) \nonumber \\
& [ ( \bfr \cdot \nabla_{\bf R} ) \bn({\bf R}) \cdot \bhu_{1}] [ ( \bfr \cdot \nabla_{\bf R} ) \bn({\bf R}) \cdot \bhu_{2}] c^{(2)} ( \bfr  , \bhu_{1}, \bhu_{2} )
\end{align}
with $\bfr$ the centre-of-mass distance between the particles and $\dot{f}$ denoting a derivative of the ODF with respect to its argument. The linear gradient term is a torque-field contribution
\begin{align}
\frac{\Delta F_{t}}{k_{B}T} & = \frac{\rho^{2}}{2} \iint d {\bf R} d \bfr  \iint d \bhu_{1} d \bhu_{2} f (\bn ({\bf R}) \cdot \bhu_{1}) \dot{f} ((\bn ({\bf R}) \cdot \bhu_{2}) \nonumber \\
& [ ( \bfr  \cdot \nabla_{\bf R} ) \bn({\bf R}) \cdot \bhu_{2}] c^{(2)} ( \bfr  , \bhu_{1}, \bhu_{2} )
\end{align}
which favors director field distortions depending on the particular choice of particle interaction (chiral or non-chiral).  The latter enters via the pair correlation functions, which is generally an unknown function of the density. In the Onsager theory,  $c^{(2)}$ simplifies to the Mayer function, $\exp[ -U/k_{B}T] -1$, of the particle potential $U$ and is independent of particle density \cite{allenevans,onsager}.   Let us consider a weak twist deformation of a homogeneous director field pointing along the $x$-axis of the  lab frame, i.e.,  $\bn ( {\bf R} ) = \be_{x} + kZ \be_{y} $ in terms of a chiral wavenumber $k$ proportional to the inverse of the cholesteric pitch.  For hard particles the spatial integral over $c^{(2)}$ reduces to an integral over the excluded-volume envelope. The twist elastic contribution per unit volume $V$ then reads
\begin{align}
 \frac{\Delta F_{2}}{Vk_{B}T} &= - \frac{k^{2}}{2} \frac{\rho^{2}}{2}  \iint d \bhu_{1} d \bhu_{2} \dot{f} (\bn \cdot \bhu_{1}) \dot{f} (\bn  \cdot \bhu_{2}) \nonumber \\ 
& \times u_{1y} u_{2y} M_{2}^{(z)} (\omega_{1}, \omega_{2}) \nonumber \\
& = \frac{k^{2}}{2} K_{2}
\label{fqua}
\end{align}
in terms of the twist elastic modulus $K_{2}$ of the nematic system. Similarly, the linear contribution translates into a free energy quantifying some effective mesoscopic twist torque exerted by the chirality of the constituent particles
\begin{align}
\frac{\Delta F_{t}}{Vk_{B}T} &= -k \frac{\rho^{2}}{2}  \iint d \bhu_{1} d \bhu_{2} f (\bn \cdot \bhu_{1}) \dot{f} (\bn \cdot \bhu_{2}) \nonumber \\
& \times u_{2y} M_{1}^{(z)} (\omega_{1}, \omega_{2}) \nonumber \\
& = k K_{t}
\label{flin}
\end{align}
Both  {\em microscopic} expressions depend on a weighted excluded volume defined as
\begin{align}
M_{n}^{( l )}  (\omega_{1}, \omega_{2}) &= \int d  \bfr_{cc} (\bfr_{cc} \cdot \be_{l} )^{n} 
\label{mm}
\end{align}
with $l$ denoting one of the Cartesian directions in the lab frame. The integral runs over the excluded volume manifold for which we invoke the parameterization expounded in Section II, i.e., $\int d \bfr_{cc} \rightarrow \prod_{i=1}^{3}  \int_{-1}^{1} dt_{i} |{\bf J}|$. Note that the zeroth moment $M_{0}^{(l)}$ recovers the total helix excluded volume $ v_{\rm excl} (\omega_{1}, \omega_{2})$. With the help of the parameterization \eq{pm} it is possible to obtain the relevant kernels for the twist parameters $K_{t}$ and $K_{2}$. In the uniaxial nematic phase the expressions simplify considerably upon replacing $M_{n}^{( \alpha )}$ by a double isotropic average over the internal angles  $\langle \langle M_{n}^{( \alpha )} \rangle \rangle_{\psi}$. This leads to more manageable expressions from which we retain only the leading order contributions in the curvature $q \delta$
\begin{align}
\frac{\langle \langle M_{1}^{(z)} \rangle \rangle_{\psi}}{2 L^{2} D  | \cos \gamma | } &\simeq -\frac{\delta^{2}q}{2}  v_{z} \nonumber \\
\frac{\langle \langle M_{2}^{(z)} \rangle \rangle _{\psi}}{2 L^{2}D| \sin \gamma |}  & \simeq    \frac{L^{2}}{12} \left ( 1 + \frac{1}{16} (q \delta)^{2} \right ) (u_{1z}^{2} + u_{2z}^{2})  
\label{m1m2}
\end{align} 
We immediately deduce that the entropic torque $K_{t}$ is zero for straight, achiral rods $(q=0)$ as expected. The  polar and azimuthal angular dependency can be rendered explicit by substituting $\bhu = \{ \cos \theta , \sin \theta \sin \varphi , \sin \theta \cos \varphi \} $. 

The equilibrium value  $k^{\ast}$ for the cholesteric wavevector simply follows from minimizing  the total free energy change $\Delta F = \Delta F_{t} + \Delta F_{2}$ with respect to $k$ and reflects a balance between torques generated by the microscopic twist and the elastic resistance.  
\beq
k^{\ast} = -\frac{K_{t}}{K_{2}}
\eeq
In this work we employ a simple Gaussian representation of the ODF to obtain analytically tractable scaling expression for strong (local) nematic order. The Gaussian trial function has been introduced by Odijk \cite{odijkoverview,vroege92} and takes the following normalized form 
\beq
f_{G}(\theta) \sim \frac{\alpha}{4 \pi} e^{-\frac{1}{2} \alpha \theta^{2}}   
 \label{ogauss}
 \eeq
with $ 0< \theta < \frac{\pi}{2}$ complemented with its mirror image for the interval $\frac{\pi}{2} < \theta < \pi$.
The asymptotic result for the twist elastic constant for straight hard rods has been derived by Odijk \cite{odijkelastic}  and scales linearly with particle concentration
\beq
\frac{K_{2}D }{k_{B}T} \sim \frac{7 c_{h}}{24 \pi} 
\label{k2gauss}
\eeq
The torque-field contribution can be quantified by taking the leading order term for small polar angles $\theta \ll1$
\begin{align}
\frac{K_{t}D^{2} }{k_{B}T} &\sim  \frac{c_{h}^{2}}{2}\left ( \frac{\delta}{L} \right )^{2} qD  \alpha  \langle \langle \gamma^{-1} (\theta_{2}^{2} -\theta_{1} \theta_{2} \cos \Delta \varphi ) \rangle \rangle_{G} \nonumber \\ 
& \propto c_{h}^{3}
\end{align}
with $\gamma \sim (\theta_{1}^{2} + \theta_{2}^{2} - 2\theta_{1}\theta_{2} \cos \Delta \varphi )^{1/2}$. The brackets denote a double Gaussian angular average using \eq{ogauss}. Although no analytical solution is available,  the scaling $ \langle \langle \gamma^{-1} (\theta_{2}^{2} -\theta_{1} \theta_{2} \cos \Delta \varphi ) \rangle \rangle_{G} \sim \alpha_{0} \alpha^{-1/2}$ is easily ascertained (with $ \alpha_{0}$ some numerical prefactor). Within the Gaussian approximation, minimizing the nematic free energy with $\alpha$ produces an analytic solution \cite{vroege92}, namely $\alpha \sim 4c_{h}^{2}/\pi$.  It subsequently follows that the concentration dependence of the torque-field is of cubic order. The inverse cholesteric pitch, then,  scales quadratically with concentration. In the limit of  asymptotically strong alignment $(c_{h} \gg 1)$ it takes the following explicit form
\beq
k^{\ast}  \sim -q \frac{24 \pi^{\frac{1}{2}}}{7}   \alpha_{0} \left ( \frac{\delta}{L} \right )^{2}  c_{h}^{2}
\label{pitch}
\eeq
with $\alpha_{0} \approx 0.886$. Several conclusions can be drawn from this result.  As expected, the cholesteric pitch length $2 \pi / k^{\ast}$ decreases with concentration but the quadratic scaling differs from the linear one predicted for long-ranged chiral interaction based on a pseudoscalar potential \cite{odijkchiral, wensinkjacksonJCP2009,wensinkJPCM2011}. This suggests that purely steric helicity leads to a much harsher microscopic twist compared to those generated by long-ranged dispersion forces. Evidence supporting the notion that a reduction of the range over which chiral intermolecular forces are transmitted may render the pitch more sensitive to concentration  can be found in measurements on {\em fd} virus suspensions where a superlinear trend is observed at high ionic strength and a sublinear one at low ionic strength \cite{grelet-fraden_chol}. In addition, numerical results for hard helices with finite aspect ratio also suggest a superlinear scaling of the pitch as a function of concentration with a marked steepening observed for decreasing helix radii \cite{dussiJCP2015}.

We emphasize that the concentration scaling in \eq{pitch}  is in  accordance with Pelcovits' prediction for hard corkscrews based on Maier-Saupe theory \cite{pelcovits1996}. We also note that the pitch is more sensitive to a change of the helix diameter $\delta$ than to  changing the molecular helicity $q$. Following Refs. \cite{odijkchiral,pelcovits1996} we may apply a rescaling recipe to gauge the pitch versus concentration for semiflexible helices. The mapping, due to Odijk \cite{odijkchiral}, consists of replacing the bare length by the deflection length $\lambda$ of the semiflexible helix:  $L \rightarrow \lambda $,  and the bare number density by the 
number density $\rho_{\lambda}$ of effective segments; $ \rho \rightarrow \rho_{\lambda}$. Using the scaling expressions $\lambda \sim (P^{1/2} D \rho)^{-2/3}$ and $\rho_{\lambda} \sim \frac{\pi}{4}(P^{2}D)^{2/3} \rho^{5/3}$ (with $P$ the persistence length of the polymer) in  \eq{pitch} leaves the concentration scaling unaltered, namely $k^{\ast} \sim -q (\delta D P \rho)^{2}$, in agreement with Ref. \onlinecite{pelcovits1996}. The power law scaling agreement with experimental observations in semiflexible polymers such as PBLG  \cite{dupreduke75}, cellulose nanofibers \cite{miller-cellulose,schutz-cellulose2015} and filamentous {\em fd} \cite{grelet-fraden_chol}.  
We finally observe that the handedness of the cholesteric structure is opposite to that of the individual helices. This is in agreement with the numerical results of Dussi et al. \cite{dussiJCP2015} for hard helices with finite aspect-ratio and small curvature $q \delta$. Particles with more outspoken helical shape may exhibit helix inversions where the handedness of the cholesteric phase changes sign  upon  variation of the microscopic helicity of the particles \cite{frezzaPCCP2014,wensinkJPCM2011} or the thermodynamic state of the system\cite{osipovPRE2000,wensinkEPL2014,belliPRE2014,dussiJCP2015}.  No such inversions are predicted for the weakly curled rods considered here.

\section{Twist-bend structure}

An alternative manifestation of helical order is a so-called twist-bend (TB) phase whereby the director field follows an oblique helicoid with the local director maintaining a constant deflection angle $\xi < \frac{\pi}{2}$ with respect to the helical axis \cite{borschNC2013}. Note that in the cholesteric phase the local director always has a right-angle tilt $\xi = \pi/2$ (see \fig{fig1}). The TB structure has been predicted theoretically \cite{dozovEPL2001} and numerically \cite{memmer2002}, both for {\em achiral} bent-core mesogens. Experimental evidence for this type of self-assembly was recently reported in \cite{borschNC2013}.  Although most bent-core molecules are achiral, a helix  with a finite diameter $\delta$ and long microscopic pitch adopts an effective bent-core shape and thus may qualify for TB order.
The director field of a chiral twist-bend phase obeys the following form
\beq
\bn({\bf R}) = \{ \cos \xi , \sin \xi \cos k X, \sin \xi \sin k X \} 
\eeq
int terms of a deflection angle $0< \xi < \frac{\pi}{2}$ and helical wavenumber $k$. Contrary to the cholesteric, the pitch of the TB phase can attain values comparable to the particle size, i.e., $kL \sim {\mathcal O} (1) $ \cite{borschNC2013}.  However, for  the weakly curved rods considered here it seems reasonable to assume the mesoscopic pitch to be compatible with the microscopic one and focus on the weak distortion limit $\xi kL \ll 1$. Hence we approximate 
\beq
\bn({\bf R}) \simeq \be_{x} + \xi \be_{y} + \xi k X \bhe_{z} 
\eeq
Inserting this into \eq{fqua} and \eq{flin} yields the distortion free energy associated with the TB director field
\beq
\frac{\Delta F}{Vk_{B}T} = \xi k K_{TB} + \frac{1}{2} (\xi k )^{2} K_{3}
\label{ftb}
\eeq
with $K_{3}$ the bend elastic constant of which the microscopic definition reads \cite{allenevans}
\beq
K_{3} =  -\frac{\rho^{2}}{2}  \iint d \bhu_{1} d \bhu_{2} \dot{f} (\bn \cdot \bhu_{1}) \dot{f} (\bn  \cdot \bhu_{2}) u_{1z} u_{2z} M_{2}^{(x)} 
\eeq
The linear-gradient contribution embodies a TB torque given by
\beq
K_{TB} =  -\frac{\rho^{2}}{2}  \iint d \bhu_{1} d \bhu_{2} f (\bn \cdot \bhu_{1}) \dot{f} (\bn  \cdot \bhu_{2}) u_{2z} M_{1}^{(x)} 
\eeq
The kernels $M$ are represented by moments of the particle excluded volume (\eq{mm})  which depend explicitly on particle orientation.  The dominant terms for weakly curled slender rods are as follows
\begin{align}
\frac{M_{1}^{(x)}}{2 L^{2} D  | \cos \gamma |}  &\simeq \frac{\delta^{2} q}{2}  v_{x} (\cos \Delta \psi -1 ) \nonumber \\
\frac{M_{2}^{(x)}}{2 L^{2} D  | \sin \gamma |}  &\simeq  \frac{L^{2}}{12} \left ( 1 + \frac{1}{16} (q \delta)^{2} \right ) (u_{1x}^{2} + u_{2x}^{2})  
\end{align} 
The effective twist-bend force $T_{TB}$ becomes zero in the uniaxial nematic phase since $M_{1}^{(x)}=0$ for a random distributions of azimuthal angle $\varphi$.  No stable twist-bend order can therefore occur in a uniaxial nematic environment.  In the biaxial phase, however,  both $K_{TB}$  and the bend elastic modulus will be non-zero and depend on the precise form of the ODF which is now a function of three angles $\{ \theta, \varphi, \psi \} $. Its density scaling can be gleaned from
\begin{align}
\frac{K_{TB}D^{2}}{k_{B}T}  \sim -c_{h}^{2} \alpha \langle \theta^{2} \rangle  \sim -c_{h}^{2}
\end{align}
A scaling expression for the bend elastic constant for straight hard rods in the Gaussian approximation is given by \cite{odijkelastic}
\beq
\frac{K_{3}D^{2}}{k_{B}T} \sim   \frac{4 }{3 \pi^{2}}   c_{h}^{3}
\eeq
The strength of  TB order can be quantified by the effective helical wavenumber $\xi k$ which upon minimizing \eq{ftb} yields an inverse proportionality with concentration, namely  $\xi k \sim 1/c_{h}$. The total free energy difference $\Delta F$ between the TB and untwisted nematic then turns out
\beq
-\frac{\Delta F}{Vk_{B}T}  \sim  \frac{K_{TB}^{2}}{K_{3}} \sim c_{h} \hspace{0.2cm} {\rm (TB)}
\eeq
whereas for the cholesteric  a much steeper density scaling is found
\beq
-\frac{\Delta F}{Vk_{B}T}  \sim \frac{K_{t}^{2}}{K_{2}} \sim c_{h}^{4}  \hspace{0.2cm} {\rm (CHOL)}
\eeq
From this we conclude that the cholesteric phase is likely be the preferred chiral mesostructure in dense systems of chiral rods in the weak curvature limit $q \delta \ll 1$.  This is in line with the numerical study of Ref. \cite{dussiJCP2015} where no evidence of stable TB order was found even for particles with a strong helical shape.

\section{Effect of local polarity}

In this section we shall take a closer look at the consequences for the cholesteric pitch  when the helices are polar, that is, when the free energy is no longer invariant upon flipping the main particle direction $\bhu \rightarrow -\bhu$. Let us propose a polarity contribution to the excess nematic free energy of the following basic form
\begin{align}
\frac{F_{p}^{{\rm excess}}}{Vk_{B}T} &\sim -\rho^{2} \tilde{u}_{p}  v_{0} \iint d \bhu_{1}d \bhu_{2} f(\bhu_{1} \cdot \bn ) f(\bhu_{2} \cdot \bn ) ( \bhu_{1} \cdot \bhu_{2}) \nonumber \\
& + \frac{k^{2}}{2}  K_{2}^{P}  
\label{fpex}
 \end{align}
with amplitude $\tilde{u}_{p} $ positive but small so that $1/\tilde{u}_{p}$ may be identified with an effective temperature. The above expression follows from applying a high-temperature expansion of the second virial terms where the potential energy consists of a chiral hard-core term and a soft perturbative polar potential  \cite{wensink_trizacPRE2014} of the form $g(\Delta r) (\bhu_{1} \cdot \bhu_{2})$.  The radial contribution $g$ depending on the inter-helix distance $\Delta r$ is assumed short-ranged of the order of the helix diameter $D$. In the fluid phase, its precise form is immaterial as only its spatial average $v_{n}  = \int d \Delta \bfr (\Delta z)^{n} g(\Delta r) \sim {\mathcal O}(D^{3+n})$ features in the prefactor $\tilde{u}_{p}$. The first contribution is a bulk one whereas the second corrects for the helical director. This terms involves an additional elastic contribution reflecting the incompatibility of polar alignment with a twist deformation of the director field. The microscopic expression for the elastic modulus imposed by the polar interactions reads
\beq
K_{2}^{(P)} \sim \rho^{2} \tilde{u}_{p}v_{2}  \iint d \bhu_{1} d \bhu_{2} \dot{f} (\bn \cdot \bhu_{1}) \dot{f} (\bn \cdot \bhu_{2}) u_{1y} u_{2y} (\bhu_{1} \cdot \bhu_{2})  \nonumber \\
\label{k2p}
\eeq 
Similar to the previous section, all angular averages featuring above will be calculated using a Gaussian trial form which we must now suitably adapt to incorporate the polar symmetry of the ODF. The simplest generalization, valid for strongly aligned systems, involves linking the Gaussian ODF to a two-state spin vector pointing parallel or anti-parallel to the nematic director 
\beq
f_{G}(\theta) \sim \frac{\alpha}{4 \pi} (2 \cosh \beta)^{-1} 
 \begin{cases} e^{-\frac{1}{2} \alpha \theta^{2}} e^{\beta} &\mbox{if } 0< \theta < \frac{\pi}{2} \\ 
 e^{-\frac{1}{2} \alpha (\pi - \theta)^{2}} e^{-\beta} & \mbox{if } \frac{\pi}{2} < \theta < \pi \end{cases} 
 \eeq
where the additional variational parameter $\beta$ must be determined self-consistently. The case $\beta \rightarrow \infty$ corresponds to the situation considered in the previous section where it was tacitly assumed that the main helix vectors all point along the local director. The other extreme case $\beta =0$ produces an apolar  state with helices pointing randomly parallel or anti-parallel to the local director. The next step is to determine the free energy induced by a spontaneous polarization of the chiral `spins'. Apart from the enthalpic term in \eq{fpex}  there is an entropic contribution associated with the polar ODF (first term in \eq{f0}).  Both can be worked out analytically in a straightforward manner.  Retaining only terms depending on $\beta$
leads to an expression that is very similar to the mean-field free energy of the one-dimensional Ising model \cite{Hill}, namely
\begin{align}
\frac{F_{p}}{Nk_{B}T} & \sim \beta \tanh \beta - \ln (\cosh \beta) - c_{h} \tilde{u}_{p} \tanh^{2} \beta \nonumber \\
&  \sim \left (\frac{1}{2} - c_{h} \tilde{u}_{p}   \right ) \beta^{2}, \hspace{0.5cm} \beta \ll 1
\end{align}
where $ T_{c} = (c_{h} \tilde{u}_{p})^{-1} = 2$ might be identified with an inverse Curie temperature separating an apolar (``paramagnetic") regime with $\beta=0$ at $T_{c} >2$  from a polar (``ferromagnetic") one $\beta >0$ at low temperatures $T_{c} < 2$. In the latter situation, the variation of polarity with temperature $\beta (c_{h} \tilde{u}_{p})$ is easily quantified numerically from minimizing the free energy $\partial F_{p} / \partial \beta =0 $. 

\begin{figure}
\begin{center}
\includegraphics[width= \columnwidth]{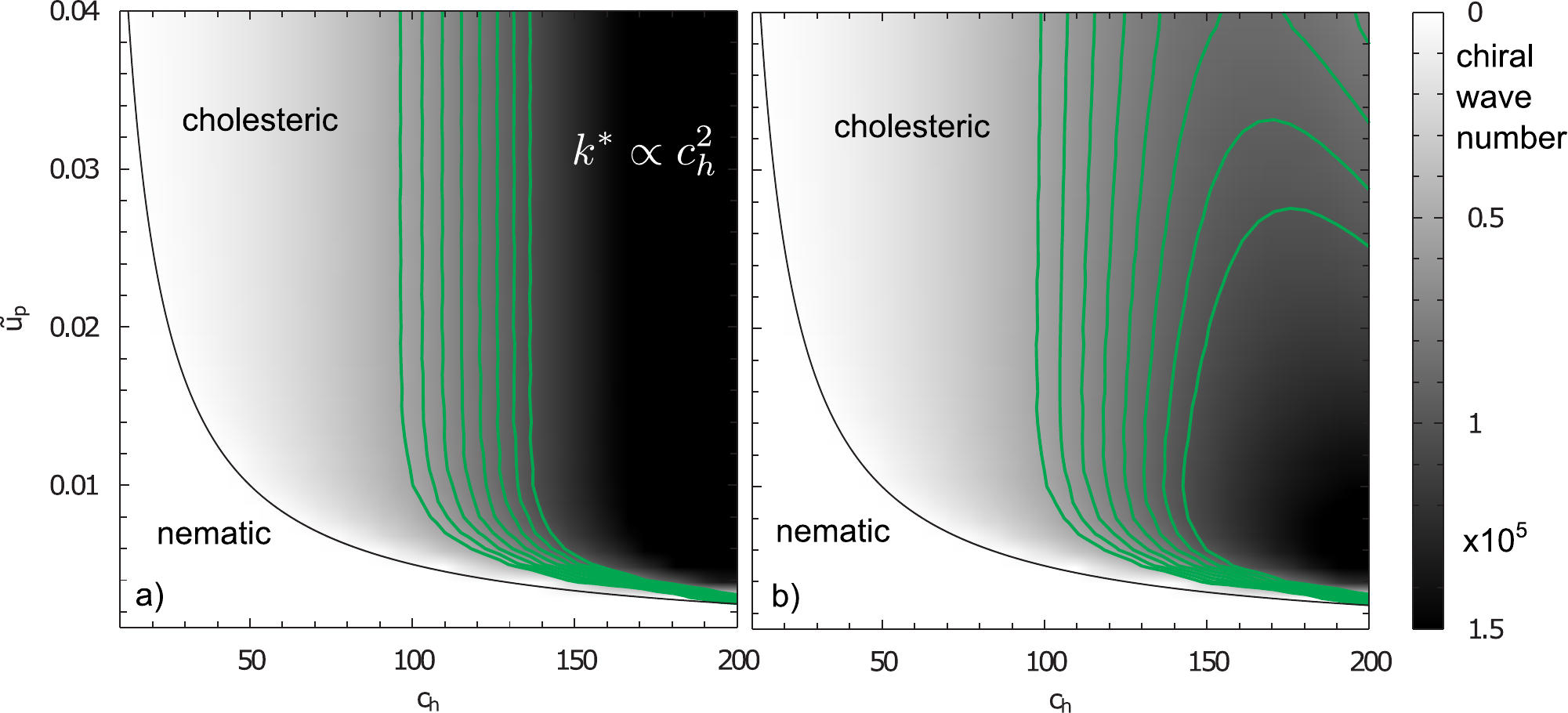}
\caption{ \label{fig3} Contour map showing the cholesteric wave number $|k^{\ast}|$ (\eq{ptemp}) versus concentration $c_{h}$ and polarity $\tilde{u}_{p}$. The values have been normalized in units $ \left ( \frac{\delta}{L} \right )^{2} |q|$. The solid black line represents the critical  curve ($c_{h}\tilde{u}_{p} =2 $) separating an apolar nematic state ($\beta =0$) from a polar cholesteric one ($\beta >0$).  Results are for (a) $D/L \downarrow 0$ and (b) $D/L = 0.002$.}
\end{center}
\end{figure}

It is now straightforward to assess the implication of local polarity on  the torque-field and twist-elastic contributions. The main twist elastic modulus \eq{fqua} turns out to be invariant with  respect to `spin' flips $\theta_{i} \rightarrow  \theta_{i} \pm \pi$ and thus remains unaffected by any variation of the polarity  $\beta$.
The polar contribution, of course, must depend explicitly on degree of polar order. Working out the Gaussian averages in \eq{k2p} yields for the leading order contribution
\begin{align}
\frac{K_{2}^{(P)}D}{k_{B}T}  \sim \tilde{u}_{p} \left ( \frac{4}{\pi} \frac{D}{L} \right )^{4} c_{h}^{6} \tanh^{2} \beta \nonumber \\
\end{align}  
We have ignored a trivial geometric constant associated with the weighted spatial average $v_{2}$.  The steep increase with concentration is off-set by the smallness of the prefactor $\tilde{u}_{p}(D/L)^{4} \ll 1$ so that the polar contribution to the elastic modulus is expected to be of the same order as the hard-core counterpart \eq{k2gauss}. Most crucially, the torque-field constant also turns out to be modified by the polar distribution.  Consequently, the cholesteric pitch \eq{pitch}  is compounded with  the local polarity of the helices in the following way 
\beq
k^{\ast}  \sim -q \frac{ \frac{24  \pi^{\frac{1}{2}}}{7}  \alpha_{0} \left ( \frac{\delta}{L} \right )^{2}  c_{h}^{2}  \tanh^{2} \beta } { 1 + \tilde{u}_{p} \frac{24 \pi}{7} \left ( \frac{4}{\pi} \frac{D}{L}  \right )^{4} c_{h}^{5} \tanh^{2} \beta }
\label{ptemp}
\eeq
The predictions for the pitch have been compiled in \fig{fig3}.  At high temperatures $\tilde{u}_{c} < 2/c_{h}$ there is no net polarity and the pitch is infinite (nematic order).  Cholestericity only sets in upon crossing the critical line and its increase with concentration is highly nonlinear. At low temperatures the pitch decreases quadratically with concentration as per \eq{pitch}.    For finite but small values $D/L$  (\fig{fig3}b) marked non-monotonic trends induced by $K_{2}^{(P)}$ are observed both upon varying concentration and temperature.  This type of behaviour stems from the fact that although both the torque-field and  the elastic resistance increase with local polar order, their scaling with concentration is different. These results clearly  demonstrate the important role of temperature in tuning the cholesteric pitch indirectly via the polarization along the local director. We emphasize that due to the helical symmetry of the director field the cholesteric state remains {\em globally } apolar. We also remark that the rich, non-monotonic behaviour is observed for weak polar amplitudes $\tilde{u}_{p} \ll 1$ corresponding to high effective temperatures where the perturbation approach should be valid  and the aligning potential of mean force between helix pairs should be dominated by their hard cores.   This falls within the realm of lyotropic systems, such as stiff biomacromolecules in solution, where particle density is the chief thermodynamic parameter (cf. \fig{fig1}) while temperature plays a secondary role. 
    
\section{Discussion and conclusion}

Using Onsager theory we have investigated the pitch of chiral nematic phases of hard helical rods in the weak deformation limit where the helix shape only slightly departs from a straight rod. We have demonstrated that the cholesteric order represents the dominant chiral mesostructure, at least  in the Onsager limit of infinite particle anisotropy. Analytical prediction for the cholesteric pitch have been proposed from assuming strong alignment along the local director and invoking Gaussian trial functions to describe the local angular distribution. The strength and sense of the cholesteric structure is always compatible with the microscopic pitch of while its magnitude decreases quadratically with concentration, in line with previous theoretical predictions. Comparing theoretical prediction with experimental measurements of the pitch in cholesteric materials (see for example Refs. \cite{schutz-cellulose2015, grelet-fraden_chol, strey-dna}) points to a strong correlation between the range through which chiral intermolecular forces are transmitted and the sensitivity of the pitch with concentration.   Accurate measurements of the concentration dependence of the pitch of lyotropic chiral materials may therefore give valuable information about whether chiral intermolecular torques are mediated via short-ranged steric forces or via long-ranged forces of e.g. electrostatic nature. In this work, we have also addressed the impact of local polarity on the pitch, hitherto underexposed in microscopic theoretical treatments,  and reveal a rich scenario where the pitch is non-trivially coupled to the local polarity of the helices. These effects might be further explored for chiral biomacromolecules. Owing to their intricate surface microstructure, many biological filaments such as actin and  {\em fd} virus possess structurally different tip ends which give rise to a distinct head-tail asymmetry  rendering the interaction between two such objects intrinsically polar.   Our results suggest that surface modification of the helix ends (using e.g. bio-engineering methods in the case of {\em fd} \cite{grelet-fraden_chol,barrySM2009}) could be a promising route to controlling the chiral mesostructure of nanohelical assemblies by subtle variations of the temperature.

\acknowledgements
L.M.A.  acknowledges the {\em Consejo Nacional de Ciencia y Tecnolog\'{i}a}, Mexico  for financial support. The authors are grateful to S. Ruzicka for helpful discussions.

\bibliography{refs}

\begin{thebibliography}{49}%
\makeatletter
\providecommand \@ifxundefined [1]{%
 \@ifx{#1\undefined}
}%
\providecommand \@ifnum [1]{%
 \ifnum #1\expandafter \@firstoftwo
 \else \expandafter \@secondoftwo
 \fi
}%
\providecommand \@ifx [1]{%
 \ifx #1\expandafter \@firstoftwo
 \else \expandafter \@secondoftwo
 \fi
}%
\providecommand \natexlab [1]{#1}%
\providecommand \enquote  [1]{``#1''}%
\providecommand \bibnamefont  [1]{#1}%
\providecommand \bibfnamefont [1]{#1}%
\providecommand \citenamefont [1]{#1}%
\providecommand \href@noop [0]{\@secondoftwo}%
\providecommand \href [0]{\begingroup \@sanitize@url \@href}%
\providecommand \@href[1]{\@@startlink{#1}\@@href}%
\providecommand \@@href[1]{\endgroup#1\@@endlink}%
\providecommand \@sanitize@url [0]{\catcode `\\12\catcode `\$12\catcode
  `\&12\catcode `\#12\catcode `\^12\catcode `\_12\catcode `\%12\relax}%
\providecommand \@@startlink[1]{}%
\providecommand \@@endlink[0]{}%
\providecommand \url  [0]{\begingroup\@sanitize@url \@url }%
\providecommand \@url [1]{\endgroup\@href {#1}{\urlprefix }}%
\providecommand \urlprefix  [0]{URL }%
\providecommand \Eprint [0]{\href }%
\providecommand \doibase [0]{http://dx.doi.org/}%
\providecommand \selectlanguage [0]{\@gobble}%
\providecommand \bibinfo  [0]{\@secondoftwo}%
\providecommand \bibfield  [0]{\@secondoftwo}%
\providecommand \translation [1]{[#1]}%
\providecommand \BibitemOpen [0]{}%
\providecommand \bibitemStop [0]{}%
\providecommand \bibitemNoStop [0]{.\EOS\space}%
\providecommand \EOS [0]{\spacefactor3000\relax}%
\providecommand \BibitemShut  [1]{\csname bibitem#1\endcsname}%
\let\auto@bib@innerbib\@empty
\bibitem [{\citenamefont {de~Gennes}\ and\ \citenamefont
  {Prost}(1993)}]{gennes-prost}%
  \BibitemOpen
  \bibfield  {author} {\bibinfo {author} {\bibfnamefont {P.~G.}\ \bibnamefont
  {de~Gennes}}\ and\ \bibinfo {author} {\bibfnamefont {J.}~\bibnamefont
  {Prost}},\ }\href@noop {} {\emph {\bibinfo {title} {The Physics of Liquid
  Crystals}}}\ (\bibinfo  {publisher} {Clarendon Press},\ \bibinfo {address}
  {Oxford},\ \bibinfo {year} {1993})\BibitemShut {NoStop}%
\bibitem [{\citenamefont {Jakli}(2013)}]{jaklibentcore2013}%
  \BibitemOpen
  \bibfield  {author} {\bibinfo {author} {\bibfnamefont {A.}~\bibnamefont
  {Jakli}},\ }\href@noop {} {\bibfield  {journal} {\bibinfo  {journal} {{Liq.
  Cryst.}}\ }\textbf {\bibinfo {volume} {{1}}},\ \bibinfo {pages} {{65}}
  (\bibinfo {year} {{2013}})}\BibitemShut {NoStop}%
\bibitem [{\citenamefont {Claessens}\ \emph {et~al.}(2008)\citenamefont
  {Claessens}, \citenamefont {Semmrich}, \citenamefont {Ramos},\ and\
  \citenamefont {Bausch}}]{claessensPNAS2008}%
  \BibitemOpen
  \bibfield  {author} {\bibinfo {author} {\bibfnamefont {M.~M. A.~E.}\
  \bibnamefont {Claessens}}, \bibinfo {author} {\bibfnamefont {C.}~\bibnamefont
  {Semmrich}}, \bibinfo {author} {\bibfnamefont {L.}~\bibnamefont {Ramos}}, \
  and\ \bibinfo {author} {\bibfnamefont {A.~R.}\ \bibnamefont {Bausch}},\
  }\href@noop {} {\bibfield  {journal} {\bibinfo  {journal} {{Proc. Nat. Acad.
  Sci.}}\ }\textbf {\bibinfo {volume} {{105}}},\ \bibinfo {pages} {{8819}}
  (\bibinfo {year} {{2008}})}\BibitemShut {NoStop}%
\bibitem [{\citenamefont {Huang}\ \emph {et~al.}(2012)\citenamefont {Huang},
  \citenamefont {Ehrhardt},\ and\ \citenamefont {Shaevitz}}]{huang2012}%
  \BibitemOpen
  \bibfield  {author} {\bibinfo {author} {\bibfnamefont {K.~C.}\ \bibnamefont
  {Huang}}, \bibinfo {author} {\bibfnamefont {D.~W.}\ \bibnamefont {Ehrhardt}},
  \ and\ \bibinfo {author} {\bibfnamefont {J.~W.}\ \bibnamefont {Shaevitz}},\
  }\href@noop {} {\bibfield  {journal} {\bibinfo  {journal} {{Curr. Opin.
  Microbiol.}}\ }\textbf {\bibinfo {volume} {{15}}},\ \bibinfo {pages} {{707}}
  (\bibinfo {year} {{2012}})}\BibitemShut {NoStop}%
\bibitem [{\citenamefont {Livolant}\ and\ \citenamefont
  {Leforestier}(1996)}]{livolantDNAoverview}%
  \BibitemOpen
  \bibfield  {author} {\bibinfo {author} {\bibfnamefont {F.}~\bibnamefont
  {Livolant}}\ and\ \bibinfo {author} {\bibfnamefont {A.}~\bibnamefont
  {Leforestier}},\ }\href@noop {} {\bibfield  {journal} {\bibinfo  {journal}
  {{Prog. Polym. Sci.}}\ }\textbf {\bibinfo {volume} {{21}}},\ \bibinfo {pages}
  {{1115}} (\bibinfo {year} {{1996}})}\BibitemShut {NoStop}%
\bibitem [{\citenamefont {Kornyshev}\ \emph {et~al.}(2007)\citenamefont
  {Kornyshev}, \citenamefont {Lee}, \citenamefont {Leikin},\ and\ \citenamefont
  {Wynveen}}]{kornyshevRMP2007}%
  \BibitemOpen
  \bibfield  {author} {\bibinfo {author} {\bibfnamefont {A.~A.}\ \bibnamefont
  {Kornyshev}}, \bibinfo {author} {\bibfnamefont {D.~J.}\ \bibnamefont {Lee}},
  \bibinfo {author} {\bibfnamefont {S.}~\bibnamefont {Leikin}}, \ and\ \bibinfo
  {author} {\bibfnamefont {A.}~\bibnamefont {Wynveen}},\ }\href@noop {}
  {\bibfield  {journal} {\bibinfo  {journal} {{Rev. Mod. Phys.}}\ }\textbf
  {\bibinfo {volume} {{79}}},\ \bibinfo {pages} {{943}} (\bibinfo {year}
  {{2007}})}\BibitemShut {NoStop}%
\bibitem [{\citenamefont {Dogic}\ and\ \citenamefont
  {Fraden}(2000)}]{dogicfradenLA2000}%
  \BibitemOpen
  \bibfield  {author} {\bibinfo {author} {\bibfnamefont {Z.}~\bibnamefont
  {Dogic}}\ and\ \bibinfo {author} {\bibfnamefont {S.}~\bibnamefont {Fraden}},\
  }\href@noop {} {\bibfield  {journal} {\bibinfo  {journal} {{Langmuir}}\
  }\textbf {\bibinfo {volume} {{16}}},\ \bibinfo {pages} {{7820}} (\bibinfo
  {year} {{2000}})}\BibitemShut {NoStop}%
\bibitem [{\citenamefont {Barry}\ \emph {et~al.}(2006)\citenamefont {Barry},
  \citenamefont {Hensel}, \citenamefont {Dogic}, \citenamefont {Shribak},\ and\
  \citenamefont {Oldenbourg}}]{barryPRL2006}%
  \BibitemOpen
  \bibfield  {author} {\bibinfo {author} {\bibfnamefont {E.}~\bibnamefont
  {Barry}}, \bibinfo {author} {\bibfnamefont {Z.}~\bibnamefont {Hensel}},
  \bibinfo {author} {\bibfnamefont {Z.}~\bibnamefont {Dogic}}, \bibinfo
  {author} {\bibfnamefont {M.}~\bibnamefont {Shribak}}, \ and\ \bibinfo
  {author} {\bibfnamefont {R.}~\bibnamefont {Oldenbourg}},\ }\href@noop {}
  {\bibfield  {journal} {\bibinfo  {journal} {{Phys. Rev. Lett.}}\ }\textbf
  {\bibinfo {volume} {{96}}},\ \bibinfo {pages} {{018305}} (\bibinfo {year}
  {{2006}})}\BibitemShut {NoStop}%
\bibitem [{\citenamefont {Amelinckx}\ \emph {et~al.}(1994)\citenamefont
  {Amelinckx}, \citenamefont {Zhang}, \citenamefont {Bernaerts}, \citenamefont
  {Zhang}, \citenamefont {Ivanov},\ and\ \citenamefont {Nagy}}]{amelinckx}%
  \BibitemOpen
  \bibfield  {author} {\bibinfo {author} {\bibfnamefont {S.}~\bibnamefont
  {Amelinckx}}, \bibinfo {author} {\bibfnamefont {X.~B.}\ \bibnamefont
  {Zhang}}, \bibinfo {author} {\bibfnamefont {D.}~\bibnamefont {Bernaerts}},
  \bibinfo {author} {\bibfnamefont {X.~F.}\ \bibnamefont {Zhang}}, \bibinfo
  {author} {\bibfnamefont {V.}~\bibnamefont {Ivanov}}, \ and\ \bibinfo {author}
  {\bibfnamefont {J.~B.}\ \bibnamefont {Nagy}},\ }\href@noop {} {\bibfield
  {journal} {\bibinfo  {journal} {{Science}}\ }\textbf {\bibinfo {volume}
  {{265}}},\ \bibinfo {pages} {{635}} (\bibinfo {year} {{1994}})}\BibitemShut
  {NoStop}%
\bibitem [{\citenamefont {Lagerwall}\ \emph {et~al.}(2014)\citenamefont
  {Lagerwall}, \citenamefont {Schutz}, \citenamefont {Salajkova}, \citenamefont
  {Noh}, \citenamefont {Park}, \citenamefont {Scalia},\ and\ \citenamefont
  {Bergstrom}}]{lagerwall2014}%
  \BibitemOpen
  \bibfield  {author} {\bibinfo {author} {\bibfnamefont {J.~P.~F.}\
  \bibnamefont {Lagerwall}}, \bibinfo {author} {\bibfnamefont {C.}~\bibnamefont
  {Schutz}}, \bibinfo {author} {\bibfnamefont {M.}~\bibnamefont {Salajkova}},
  \bibinfo {author} {\bibfnamefont {J.}~\bibnamefont {Noh}}, \bibinfo {author}
  {\bibfnamefont {J.~H.}\ \bibnamefont {Park}}, \bibinfo {author}
  {\bibfnamefont {G.}~\bibnamefont {Scalia}}, \ and\ \bibinfo {author}
  {\bibfnamefont {L.}~\bibnamefont {Bergstrom}},\ }\href@noop {} {\bibfield
  {journal} {\bibinfo  {journal} {{NPG Asia Mater.}}\ }\textbf {\bibinfo
  {volume} {{6}}},\ \bibinfo {pages} {{e80}} (\bibinfo {year}
  {{2014}})}\BibitemShut {NoStop}%
\bibitem [{\citenamefont {Schutz}\ \emph {et~al.}(2015)\citenamefont {Schutz},
  \citenamefont {Agthe}, \citenamefont {Fall}, \citenamefont {Gordeyeva},
  \citenamefont {Guccini}, \citenamefont {Salajkova}, \citenamefont {Plivelic},
  \citenamefont {Lagerwall}, \citenamefont {Salazar-Alvarez},\ and\
  \citenamefont {Bergstrom}}]{schutz-cellulose2015}%
  \BibitemOpen
  \bibfield  {author} {\bibinfo {author} {\bibfnamefont {C.}~\bibnamefont
  {Schutz}}, \bibinfo {author} {\bibfnamefont {M.}~\bibnamefont {Agthe}},
  \bibinfo {author} {\bibfnamefont {A.~B.}\ \bibnamefont {Fall}}, \bibinfo
  {author} {\bibfnamefont {K.}~\bibnamefont {Gordeyeva}}, \bibinfo {author}
  {\bibfnamefont {V.}~\bibnamefont {Guccini}}, \bibinfo {author} {\bibfnamefont
  {M.}~\bibnamefont {Salajkova}}, \bibinfo {author} {\bibfnamefont {T.~S.}\
  \bibnamefont {Plivelic}}, \bibinfo {author} {\bibfnamefont {J.~P.~F.}\
  \bibnamefont {Lagerwall}}, \bibinfo {author} {\bibfnamefont {G.}~\bibnamefont
  {Salazar-Alvarez}}, \ and\ \bibinfo {author} {\bibfnamefont {L.}~\bibnamefont
  {Bergstrom}},\ }\href@noop {} {\bibfield  {journal} {\bibinfo  {journal}
  {{Langmuir}}\ }\textbf {\bibinfo {volume} {{31}}},\ \bibinfo {pages} {{6507}}
  (\bibinfo {year} {{2015}})}\BibitemShut {NoStop}%
\bibitem [{\citenamefont {Pham}\ \emph {et~al.}(2013)\citenamefont {Pham},
  \citenamefont {Lawrence}, \citenamefont {Lee}, \citenamefont {Grason},
  \citenamefont {Emrick},\ and\ \citenamefont {Crosby}}]{crosbyAM2013}%
  \BibitemOpen
  \bibfield  {author} {\bibinfo {author} {\bibfnamefont {J.~T.}\ \bibnamefont
  {Pham}}, \bibinfo {author} {\bibfnamefont {J.}~\bibnamefont {Lawrence}},
  \bibinfo {author} {\bibfnamefont {D.~Y.}\ \bibnamefont {Lee}}, \bibinfo
  {author} {\bibfnamefont {G.~M.}\ \bibnamefont {Grason}}, \bibinfo {author}
  {\bibfnamefont {T.}~\bibnamefont {Emrick}}, \ and\ \bibinfo {author}
  {\bibfnamefont {A.~J.}\ \bibnamefont {Crosby}},\ }\href@noop {} {\bibfield
  {journal} {\bibinfo  {journal} {{Adv. Mater.}}\ }\textbf {\bibinfo {volume}
  {{25}}},\ \bibinfo {pages} {{6703}} (\bibinfo {year} {{2013}})}\BibitemShut
  {NoStop}%
\bibitem [{\citenamefont {Harris}\ \emph {et~al.}(1997)\citenamefont {Harris},
  \citenamefont {Kamien},\ and\ \citenamefont
  {Lubensky}}]{harris-kamienPRL1997}%
  \BibitemOpen
  \bibfield  {author} {\bibinfo {author} {\bibfnamefont {A.~B.}\ \bibnamefont
  {Harris}}, \bibinfo {author} {\bibfnamefont {R.~D.}\ \bibnamefont {Kamien}},
  \ and\ \bibinfo {author} {\bibfnamefont {T.~C.}\ \bibnamefont {Lubensky}},\
  }\href@noop {} {\bibfield  {journal} {\bibinfo  {journal} {Phys. Rev. Lett.}\
  }\textbf {\bibinfo {volume} {78}},\ \bibinfo {pages} {1476} (\bibinfo {year}
  {1997})}\BibitemShut {NoStop}%
\bibitem [{\citenamefont {Emelyanenko}(2003)}]{emelyanenkoPRE2003}%
  \BibitemOpen
  \bibfield  {author} {\bibinfo {author} {\bibfnamefont {A.~V.}\ \bibnamefont
  {Emelyanenko}},\ }\href@noop {} {\bibfield  {journal} {\bibinfo  {journal}
  {{Phys. Rev. E}}\ }\textbf {\bibinfo {volume} {{67}}},\ \bibinfo {pages}
  {{031704}} (\bibinfo {year} {{2003}})}\BibitemShut {NoStop}%
\bibitem [{\citenamefont {Goossens}(1971)}]{goossens1971}%
  \BibitemOpen
  \bibfield  {author} {\bibinfo {author} {\bibfnamefont {W.~J.~A.}\
  \bibnamefont {Goossens}},\ }\href@noop {} {\bibfield  {journal} {\bibinfo
  {journal} {Mol. Cryst. Liq. Cryst.}\ }\textbf {\bibinfo {volume} {12}},\
  \bibinfo {pages} {237} (\bibinfo {year} {1971})}\BibitemShut {NoStop}%
\bibitem [{\citenamefont {Odijk}(1987)}]{odijkchiral}%
  \BibitemOpen
  \bibfield  {author} {\bibinfo {author} {\bibfnamefont {T.}~\bibnamefont
  {Odijk}},\ }\href@noop {} {\bibfield  {journal} {\bibinfo  {journal} {J.
  Phys. Chem.}\ }\textbf {\bibinfo {volume} {91}},\ \bibinfo {pages} {6060}
  (\bibinfo {year} {1987})}\BibitemShut {NoStop}%
\bibitem [{\citenamefont {Varga}\ and\ \citenamefont
  {Jackson}(2006)}]{vargachiral1}%
  \BibitemOpen
  \bibfield  {author} {\bibinfo {author} {\bibfnamefont {S.}~\bibnamefont
  {Varga}}\ and\ \bibinfo {author} {\bibfnamefont {G.}~\bibnamefont
  {Jackson}},\ }\href@noop {} {\bibfield  {journal} {\bibinfo  {journal} {Mol.
  Phys.}\ }\textbf {\bibinfo {volume} {104}},\ \bibinfo {pages} {3681}
  (\bibinfo {year} {2006})}\BibitemShut {NoStop}%
\bibitem [{\citenamefont {Wensink}(2014)}]{wensinkEPL2014}%
  \BibitemOpen
  \bibfield  {author} {\bibinfo {author} {\bibfnamefont {H.~H.}\ \bibnamefont
  {Wensink}},\ }\href@noop {} {\bibfield  {journal} {\bibinfo  {journal}
  {{EPL}}\ }\textbf {\bibinfo {volume} {{107}}},\ \bibinfo {pages} {{36001}}
  (\bibinfo {year} {{2014}})}\BibitemShut {NoStop}%
\bibitem [{\citenamefont {Wensink}\ and\ \citenamefont
  {Jackson}(2011)}]{wensinkJPCM2011}%
  \BibitemOpen
  \bibfield  {author} {\bibinfo {author} {\bibfnamefont {H.~H.}\ \bibnamefont
  {Wensink}}\ and\ \bibinfo {author} {\bibfnamefont {G.}~\bibnamefont
  {Jackson}},\ }\href@noop {} {\bibfield  {journal} {\bibinfo  {journal} {{J.
  Phys.: Condens. Matter}}\ }\textbf {\bibinfo {volume} {{23}}},\ \bibinfo
  {pages} {{194107}} (\bibinfo {year} {{2011}})}\BibitemShut {NoStop}%
\bibitem [{\citenamefont {Belli}\ \emph {et~al.}(2014)\citenamefont {Belli},
  \citenamefont {Dussi}, \citenamefont {Dijkstra},\ and\ \citenamefont {van
  Roij}}]{belliPRE2014}%
  \BibitemOpen
  \bibfield  {author} {\bibinfo {author} {\bibfnamefont {S.}~\bibnamefont
  {Belli}}, \bibinfo {author} {\bibfnamefont {S.}~\bibnamefont {Dussi}},
  \bibinfo {author} {\bibfnamefont {M.}~\bibnamefont {Dijkstra}}, \ and\
  \bibinfo {author} {\bibfnamefont {R.}~\bibnamefont {van Roij}},\ }\href@noop
  {} {\bibfield  {journal} {\bibinfo  {journal} {{Phys. Rev. E}}\ }\textbf
  {\bibinfo {volume} {{90}}},\ \bibinfo {pages} {{020503}} (\bibinfo {year}
  {{2014}})}\BibitemShut {NoStop}%
\bibitem [{\citenamefont {Kolli}\ \emph
  {et~al.}(2014{\natexlab{a}})\citenamefont {Kolli}, \citenamefont {Frezza},
  \citenamefont {Cinacchi}, \citenamefont {Ferrarini}, \citenamefont
  {Giacometti}, \citenamefont {Hudson}, \citenamefont {DeMichele},\ and\
  \citenamefont {Sciortino}}]{kolliSM2014}%
  \BibitemOpen
  \bibfield  {author} {\bibinfo {author} {\bibfnamefont {H.~B.}\ \bibnamefont
  {Kolli}}, \bibinfo {author} {\bibfnamefont {E.}~\bibnamefont {Frezza}},
  \bibinfo {author} {\bibfnamefont {G.}~\bibnamefont {Cinacchi}}, \bibinfo
  {author} {\bibfnamefont {A.}~\bibnamefont {Ferrarini}}, \bibinfo {author}
  {\bibfnamefont {A.}~\bibnamefont {Giacometti}}, \bibinfo {author}
  {\bibfnamefont {T.~S.}\ \bibnamefont {Hudson}}, \bibinfo {author}
  {\bibfnamefont {C.}~\bibnamefont {DeMichele}}, \ and\ \bibinfo {author}
  {\bibfnamefont {G.}~\bibnamefont {Sciortino}},\ }\href@noop {} {\bibfield
  {journal} {\bibinfo  {journal} {{Soft Matter}}\ }\textbf {\bibinfo {volume}
  {{10}}},\ \bibinfo {pages} {{8171}} (\bibinfo {year}
  {{2014}}{\natexlab{a}})}\BibitemShut {NoStop}%
\bibitem [{\citenamefont {Evans}(1992)}]{evansMP1992}%
  \BibitemOpen
  \bibfield  {author} {\bibinfo {author} {\bibfnamefont {G.~T.}\ \bibnamefont
  {Evans}},\ }\href@noop {} {\bibfield  {journal} {\bibinfo  {journal} {{Mol.
  Phys.}}\ }\textbf {\bibinfo {volume} {{77}}},\ \bibinfo {pages} {{969}}
  (\bibinfo {year} {{1992}})}\BibitemShut {NoStop}%
\bibitem [{\citenamefont {Pelcovits}(1996)}]{pelcovits1996}%
  \BibitemOpen
  \bibfield  {author} {\bibinfo {author} {\bibfnamefont {R.~A.}\ \bibnamefont
  {Pelcovits}},\ }\href@noop {} {\bibfield  {journal} {\bibinfo  {journal}
  {Liq. Cryst.}\ }\textbf {\bibinfo {volume} {21}},\ \bibinfo {pages} {361}
  (\bibinfo {year} {1996})}\BibitemShut {NoStop}%
\bibitem [{\citenamefont {Varga}\ and\ \citenamefont
  {Jackson}(2011)}]{vargajackson2011}%
  \BibitemOpen
  \bibfield  {author} {\bibinfo {author} {\bibfnamefont {S.}~\bibnamefont
  {Varga}}\ and\ \bibinfo {author} {\bibfnamefont {G.}~\bibnamefont
  {Jackson}},\ }\href@noop {} {\bibfield  {journal} {\bibinfo  {journal} {{Mol.
  Phys.}}\ }\textbf {\bibinfo {volume} {{109}}},\ \bibinfo {pages} {{1313}}
  (\bibinfo {year} {{2011}})}\BibitemShut {NoStop}%
\bibitem [{\citenamefont {DuPr\'{e}}\ and\ \citenamefont
  {Duke}(1975)}]{dupreduke75}%
  \BibitemOpen
  \bibfield  {author} {\bibinfo {author} {\bibfnamefont {D.~B.}\ \bibnamefont
  {DuPr\'{e}}}\ and\ \bibinfo {author} {\bibfnamefont {R.~W.}\ \bibnamefont
  {Duke}},\ }\href@noop {} {\bibfield  {journal} {\bibinfo  {journal} {J. Chem.
  Phys.}\ }\textbf {\bibinfo {volume} {63}},\ \bibinfo {pages} {143} (\bibinfo
  {year} {1975})}\BibitemShut {NoStop}%
\bibitem [{\citenamefont {Werbowyj}\ and\ \citenamefont
  {Gray}(1976)}]{gray-cullulose}%
  \BibitemOpen
  \bibfield  {author} {\bibinfo {author} {\bibfnamefont {R.~S.}\ \bibnamefont
  {Werbowyj}}\ and\ \bibinfo {author} {\bibfnamefont {D.~G.}\ \bibnamefont
  {Gray}},\ }\href@noop {} {\bibfield  {journal} {\bibinfo  {journal} {Mol.
  Cryst. Liq. Cryst.}\ }\textbf {\bibinfo {volume} {34}},\ \bibinfo {pages}
  {97} (\bibinfo {year} {1976})}\BibitemShut {NoStop}%
\bibitem [{\citenamefont {Miller}\ and\ \citenamefont
  {Donald}(2003)}]{miller-cellulose}%
  \BibitemOpen
  \bibfield  {author} {\bibinfo {author} {\bibfnamefont {A.~F.}\ \bibnamefont
  {Miller}}\ and\ \bibinfo {author} {\bibfnamefont {A.~M.}\ \bibnamefont
  {Donald}},\ }\href@noop {} {\bibfield  {journal} {\bibinfo  {journal}
  {Biomacromolecules}\ }\textbf {\bibinfo {volume} {4}},\ \bibinfo {pages}
  {510} (\bibinfo {year} {2003})}\BibitemShut {NoStop}%
\bibitem [{\citenamefont {Onsager}(1949)}]{onsager}%
  \BibitemOpen
  \bibfield  {author} {\bibinfo {author} {\bibfnamefont {L.}~\bibnamefont
  {Onsager}},\ }\href@noop {} {\bibfield  {journal} {\bibinfo  {journal} {Ann.
  N.Y. Acad. Sci.}\ }\textbf {\bibinfo {volume} {51}},\ \bibinfo {pages} {627}
  (\bibinfo {year} {1949})}\BibitemShut {NoStop}%
\bibitem [{\citenamefont {Straley}(1976)}]{straleychiral}%
  \BibitemOpen
  \bibfield  {author} {\bibinfo {author} {\bibfnamefont {J.~P.}\ \bibnamefont
  {Straley}},\ }\href@noop {} {\bibfield  {journal} {\bibinfo  {journal} {Phys.
  Rev. A}\ }\textbf {\bibinfo {volume} {14}},\ \bibinfo {pages} {1835}
  (\bibinfo {year} {1976})}\BibitemShut {NoStop}%
\bibitem [{\citenamefont {Dozov}(2001)}]{dozovEPL2001}%
  \BibitemOpen
  \bibfield  {author} {\bibinfo {author} {\bibfnamefont {I.}~\bibnamefont
  {Dozov}},\ }\href@noop {} {\bibfield  {journal} {\bibinfo  {journal} {{EPL}}\
  }\textbf {\bibinfo {volume} {{56}}},\ \bibinfo {pages} {{247}} (\bibinfo
  {year} {{2001}})}\BibitemShut {NoStop}%
\bibitem [{\citenamefont {Memmer}(2002)}]{memmer2002}%
  \BibitemOpen
  \bibfield  {author} {\bibinfo {author} {\bibfnamefont {R.}~\bibnamefont
  {Memmer}},\ }\href@noop {} {\bibfield  {journal} {\bibinfo  {journal} {Liq.
  Cryst.}\ }\textbf {\bibinfo {volume} {29}},\ \bibinfo {pages} {483} (\bibinfo
  {year} {2002})}\BibitemShut {NoStop}%
\bibitem [{\citenamefont {Borshch}\ \emph {et~al.}(2013)\citenamefont
  {Borshch}, \citenamefont {Kim}, \citenamefont {Xiang}, \citenamefont {Gao},
  \citenamefont {Jakli}, \citenamefont {Panov}, \citenamefont {Vij},
  \citenamefont {Imrie}, \citenamefont {Tamba}, \citenamefont {Mehl},\ and\
  \citenamefont {Lavrentovich}}]{borschNC2013}%
  \BibitemOpen
  \bibfield  {author} {\bibinfo {author} {\bibfnamefont {V.}~\bibnamefont
  {Borshch}}, \bibinfo {author} {\bibfnamefont {Y.~K.}\ \bibnamefont {Kim}},
  \bibinfo {author} {\bibfnamefont {J.}~\bibnamefont {Xiang}}, \bibinfo
  {author} {\bibfnamefont {M.}~\bibnamefont {Gao}}, \bibinfo {author}
  {\bibfnamefont {A.}~\bibnamefont {Jakli}}, \bibinfo {author} {\bibfnamefont
  {V.~P.}\ \bibnamefont {Panov}}, \bibinfo {author} {\bibfnamefont {J.~K.}\
  \bibnamefont {Vij}}, \bibinfo {author} {\bibfnamefont {C.~T.}\ \bibnamefont
  {Imrie}}, \bibinfo {author} {\bibfnamefont {M.~G.}\ \bibnamefont {Tamba}},
  \bibinfo {author} {\bibfnamefont {G.~H.}\ \bibnamefont {Mehl}}, \ and\
  \bibinfo {author} {\bibfnamefont {O.~D.}\ \bibnamefont {Lavrentovich}},\
  }\href@noop {} {\bibfield  {journal} {\bibinfo  {journal} {{Nat. Commun.}}\
  }\textbf {\bibinfo {volume} {{4}}},\ \bibinfo {pages} {{2635}} (\bibinfo
  {year} {{2013}})}\BibitemShut {NoStop}%
\bibitem [{\citenamefont {Kolli}\ \emph
  {et~al.}(2014{\natexlab{b}})\citenamefont {Kolli}, \citenamefont {Frezza},
  \citenamefont {Cinacchi}, \citenamefont {Ferrarini}, \citenamefont
  {Giacometti},\ and\ \citenamefont {Hudson}}]{kolliJCP2014}%
  \BibitemOpen
  \bibfield  {author} {\bibinfo {author} {\bibfnamefont {H.~B.}\ \bibnamefont
  {Kolli}}, \bibinfo {author} {\bibfnamefont {E.}~\bibnamefont {Frezza}},
  \bibinfo {author} {\bibfnamefont {G.}~\bibnamefont {Cinacchi}}, \bibinfo
  {author} {\bibfnamefont {A.}~\bibnamefont {Ferrarini}}, \bibinfo {author}
  {\bibfnamefont {A.}~\bibnamefont {Giacometti}}, \ and\ \bibinfo {author}
  {\bibfnamefont {T.~S.}\ \bibnamefont {Hudson}},\ }\href@noop {} {\bibfield
  {journal} {\bibinfo  {journal} {{J. Chem. Phys.}}\ }\textbf {\bibinfo
  {volume} {{140}}},\ \bibinfo {pages} {{081101}} (\bibinfo {year}
  {{2014}}{\natexlab{b}})}\BibitemShut {NoStop}%
\bibitem [{\citenamefont {Vroege}\ and\ \citenamefont
  {Lekkerkerker}(1992)}]{vroege92}%
  \BibitemOpen
  \bibfield  {author} {\bibinfo {author} {\bibfnamefont {G.~J.}\ \bibnamefont
  {Vroege}}\ and\ \bibinfo {author} {\bibfnamefont {H.~N.~W.}\ \bibnamefont
  {Lekkerkerker}},\ }\href@noop {} {\bibfield  {journal} {\bibinfo  {journal}
  {Rep. Prog. Phys.}\ }\textbf {\bibinfo {volume} {55}},\ \bibinfo {pages}
  {1241} (\bibinfo {year} {1992})}\BibitemShut {NoStop}%
\bibitem [{\citenamefont {Lekkerkerker}\ \emph {et~al.}(1984)\citenamefont
  {Lekkerkerker}, \citenamefont {Coulon}, \citenamefont {van~der Hagen},\ and\
  \citenamefont {Deblieck}}]{Lekkerkerker84}%
  \BibitemOpen
  \bibfield  {author} {\bibinfo {author} {\bibfnamefont {H.~N.~W.}\
  \bibnamefont {Lekkerkerker}}, \bibinfo {author} {\bibfnamefont
  {P.}~\bibnamefont {Coulon}}, \bibinfo {author} {\bibfnamefont
  {R.}~\bibnamefont {van~der Hagen}}, \ and\ \bibinfo {author} {\bibfnamefont
  {R.}~\bibnamefont {Deblieck}},\ }\href@noop {} {\bibfield  {journal}
  {\bibinfo  {journal} {J. Chem. Phys.}\ }\textbf {\bibinfo {volume} {80}},\
  \bibinfo {pages} {3427} (\bibinfo {year} {1984})}\BibitemShut {NoStop}%
\bibitem [{\citenamefont {Kayser}\ and\ \citenamefont
  {Ravech\'{e}}(1978)}]{kayser}%
  \BibitemOpen
  \bibfield  {author} {\bibinfo {author} {\bibfnamefont {R.~F.}\ \bibnamefont
  {Kayser}}\ and\ \bibinfo {author} {\bibfnamefont {H.~J.}\ \bibnamefont
  {Ravech\'{e}}},\ }\href@noop {} {\bibfield  {journal} {\bibinfo  {journal}
  {Phys. Rev. A}\ }\textbf {\bibinfo {volume} {17}},\ \bibinfo {pages} {2067}
  (\bibinfo {year} {1978})}\BibitemShut {NoStop}%
\bibitem [{\citenamefont {Frezza}\ \emph {et~al.}(2013)\citenamefont {Frezza},
  \citenamefont {Ferrarini}, \citenamefont {Kolli}, \citenamefont
  {Giacometti},\ and\ \citenamefont {Cinacchi}}]{frezzaJCP2013}%
  \BibitemOpen
  \bibfield  {author} {\bibinfo {author} {\bibfnamefont {E.}~\bibnamefont
  {Frezza}}, \bibinfo {author} {\bibfnamefont {A.}~\bibnamefont {Ferrarini}},
  \bibinfo {author} {\bibfnamefont {H.~B.}\ \bibnamefont {Kolli}}, \bibinfo
  {author} {\bibfnamefont {A.}~\bibnamefont {Giacometti}}, \ and\ \bibinfo
  {author} {\bibfnamefont {G.}~\bibnamefont {Cinacchi}},\ }\href@noop {}
  {\bibfield  {journal} {\bibinfo  {journal} {{J. Chem. Phys.}}\ }\textbf
  {\bibinfo {volume} {{138}}},\ \bibinfo {pages} {{164906}} (\bibinfo {year}
  {{2013}})}\BibitemShut {NoStop}%
\bibitem [{\citenamefont {Allen}\ \emph {et~al.}(1993)\citenamefont {Allen},
  \citenamefont {Evans}, \citenamefont {Frenkel},\ and\ \citenamefont
  {Mulder}}]{allenevans}%
  \BibitemOpen
  \bibfield  {author} {\bibinfo {author} {\bibfnamefont {M.~P.}\ \bibnamefont
  {Allen}}, \bibinfo {author} {\bibfnamefont {G.~T.}\ \bibnamefont {Evans}},
  \bibinfo {author} {\bibfnamefont {D.}~\bibnamefont {Frenkel}}, \ and\
  \bibinfo {author} {\bibfnamefont {B.~M.}\ \bibnamefont {Mulder}},\
  }\href@noop {} {\bibfield  {journal} {\bibinfo  {journal} {Adv. Chem. Phys.}\
  }\textbf {\bibinfo {volume} {86}},\ \bibinfo {pages} {1} (\bibinfo {year}
  {1993})}\BibitemShut {NoStop}%
\bibitem [{\citenamefont {Odijk}(1986{\natexlab{a}})}]{odijkoverview}%
  \BibitemOpen
  \bibfield  {author} {\bibinfo {author} {\bibfnamefont {T.}~\bibnamefont
  {Odijk}},\ }\href@noop {} {\bibfield  {journal} {\bibinfo  {journal}
  {Macromolecules}\ }\textbf {\bibinfo {volume} {19}},\ \bibinfo {pages} {2313}
  (\bibinfo {year} {1986}{\natexlab{a}})}\BibitemShut {NoStop}%
\bibitem [{\citenamefont {Odijk}(1986{\natexlab{b}})}]{odijkelastic}%
  \BibitemOpen
  \bibfield  {author} {\bibinfo {author} {\bibfnamefont {T.}~\bibnamefont
  {Odijk}},\ }\href@noop {} {\bibfield  {journal} {\bibinfo  {journal} {Liq.
  Cryst.}\ }\textbf {\bibinfo {volume} {1}},\ \bibinfo {pages} {553} (\bibinfo
  {year} {1986}{\natexlab{b}})}\BibitemShut {NoStop}%
\bibitem [{\citenamefont {Wensink}\ and\ \citenamefont
  {Jackson}(2009)}]{wensinkjacksonJCP2009}%
  \BibitemOpen
  \bibfield  {author} {\bibinfo {author} {\bibfnamefont {H.~H.}\ \bibnamefont
  {Wensink}}\ and\ \bibinfo {author} {\bibfnamefont {G.}~\bibnamefont
  {Jackson}},\ }\href@noop {} {\bibfield  {journal} {\bibinfo  {journal} {J.
  Chem. Phys.}\ }\textbf {\bibinfo {volume} {130}},\ \bibinfo {pages} {234911}
  (\bibinfo {year} {2009})}\BibitemShut {NoStop}%
\bibitem [{\citenamefont {Grelet}\ and\ \citenamefont
  {Fraden}(2003)}]{grelet-fraden_chol}%
  \BibitemOpen
  \bibfield  {author} {\bibinfo {author} {\bibfnamefont {E.}~\bibnamefont
  {Grelet}}\ and\ \bibinfo {author} {\bibfnamefont {S.}~\bibnamefont
  {Fraden}},\ }\href@noop {} {\bibfield  {journal} {\bibinfo  {journal} {Phys.
  Rev. Lett.}\ }\textbf {\bibinfo {volume} {90}},\ \bibinfo {pages} {198302}
  (\bibinfo {year} {2003})}\BibitemShut {NoStop}%
\bibitem [{\citenamefont {Dussi}\ \emph {et~al.}(2015)\citenamefont {Dussi},
  \citenamefont {Belli}, \citenamefont {van Roij},\ and\ \citenamefont
  {Dijkstra}}]{dussiJCP2015}%
  \BibitemOpen
  \bibfield  {author} {\bibinfo {author} {\bibfnamefont {S.}~\bibnamefont
  {Dussi}}, \bibinfo {author} {\bibfnamefont {S.}~\bibnamefont {Belli}},
  \bibinfo {author} {\bibfnamefont {R.}~\bibnamefont {van Roij}}, \ and\
  \bibinfo {author} {\bibfnamefont {M.}~\bibnamefont {Dijkstra}},\ }\href@noop
  {} {\bibfield  {journal} {\bibinfo  {journal} {{J. Chem. Phys.}}\ }\textbf
  {\bibinfo {volume} {{142}}},\ \bibinfo {pages} {{074905}} (\bibinfo {year}
  {{2015}})}\BibitemShut {NoStop}%
\bibitem [{\citenamefont {Frezza}\ \emph {et~al.}(2014)\citenamefont {Frezza},
  \citenamefont {Ferrarini}, \citenamefont {Kolli}, \citenamefont
  {Giacometti},\ and\ \citenamefont {Cinacchi}}]{frezzaPCCP2014}%
  \BibitemOpen
  \bibfield  {author} {\bibinfo {author} {\bibfnamefont {E.}~\bibnamefont
  {Frezza}}, \bibinfo {author} {\bibfnamefont {A.}~\bibnamefont {Ferrarini}},
  \bibinfo {author} {\bibfnamefont {H.~B.}\ \bibnamefont {Kolli}}, \bibinfo
  {author} {\bibfnamefont {A.}~\bibnamefont {Giacometti}}, \ and\ \bibinfo
  {author} {\bibfnamefont {G.}~\bibnamefont {Cinacchi}},\ }\href@noop {}
  {\bibfield  {journal} {\bibinfo  {journal} {{Phys. Chem. Chem. Phys.}}\
  }\textbf {\bibinfo {volume} {{16}}},\ \bibinfo {pages} {{16225}} (\bibinfo
  {year} {{2014}})}\BibitemShut {NoStop}%
\bibitem [{\citenamefont {Emelyanenko}\ \emph {et~al.}(2000)\citenamefont
  {Emelyanenko}, \citenamefont {Osipov},\ and\ \citenamefont
  {Dunmur}}]{osipovPRE2000}%
  \BibitemOpen
  \bibfield  {author} {\bibinfo {author} {\bibfnamefont {A.~V.}\ \bibnamefont
  {Emelyanenko}}, \bibinfo {author} {\bibfnamefont {M.~A.}\ \bibnamefont
  {Osipov}}, \ and\ \bibinfo {author} {\bibfnamefont {D.~A.}\ \bibnamefont
  {Dunmur}},\ }\href@noop {} {\bibfield  {journal} {\bibinfo  {journal} {Phys.
  Rev. E}\ }\textbf {\bibinfo {volume} {62}},\ \bibinfo {pages} {2340}
  (\bibinfo {year} {2000})}\BibitemShut {NoStop}%
\bibitem [{\citenamefont {Wensink}\ and\ \citenamefont
  {Trizac}(2014)}]{wensink_trizacPRE2014}%
  \BibitemOpen
  \bibfield  {author} {\bibinfo {author} {\bibfnamefont {H.~H.}\ \bibnamefont
  {Wensink}}\ and\ \bibinfo {author} {\bibfnamefont {E.}~\bibnamefont
  {Trizac}},\ }\href@noop {} {\bibfield  {journal} {\bibinfo  {journal} {{J.
  Chem. Phys.}}\ }\textbf {\bibinfo {volume} {{140}}},\ \bibinfo {pages}
  {{024901}} (\bibinfo {year} {{2014}})}\BibitemShut {NoStop}%
\bibitem [{\citenamefont {Hill}(1956)}]{Hill}%
  \BibitemOpen
  \bibfield  {author} {\bibinfo {author} {\bibfnamefont {T.~L.}\ \bibnamefont
  {Hill}},\ }\href@noop {} {\emph {\bibinfo {title} {Statistical Mechanics}}}\
  (\bibinfo  {publisher} {McGraw-Hill},\ \bibinfo {address} {New York},\
  \bibinfo {year} {1956})\BibitemShut {NoStop}%
\bibitem [{\citenamefont {Stanley}\ \emph {et~al.}(2005)\citenamefont
  {Stanley}, \citenamefont {Hong},\ and\ \citenamefont {Strey}}]{strey-dna}%
  \BibitemOpen
  \bibfield  {author} {\bibinfo {author} {\bibfnamefont {C.~B.}\ \bibnamefont
  {Stanley}}, \bibinfo {author} {\bibfnamefont {H.}~\bibnamefont {Hong}}, \
  and\ \bibinfo {author} {\bibfnamefont {H.~H.}\ \bibnamefont {Strey}},\
  }\href@noop {} {\bibfield  {journal} {\bibinfo  {journal} {Biophys. J.}\
  }\textbf {\bibinfo {volume} {89}},\ \bibinfo {pages} {2552} (\bibinfo {year}
  {2005})}\BibitemShut {NoStop}%
\bibitem [{\citenamefont {Barry}\ \emph {et~al.}(2009)\citenamefont {Barry},
  \citenamefont {Beller},\ and\ \citenamefont {Dogic}}]{barrySM2009}%
  \BibitemOpen
  \bibfield  {author} {\bibinfo {author} {\bibfnamefont {E.}~\bibnamefont
  {Barry}}, \bibinfo {author} {\bibfnamefont {D.}~\bibnamefont {Beller}}, \
  and\ \bibinfo {author} {\bibfnamefont {Z.}~\bibnamefont {Dogic}},\
  }\href@noop {} {\bibfield  {journal} {\bibinfo  {journal} {Soft Matter}\
  }\textbf {\bibinfo {volume} {5}},\ \bibinfo {pages} {2563} (\bibinfo {year}
  {2009})}\BibitemShut {NoStop}%
\end{thebibliography}%

\end{document}